\documentclass[]{iopart}
\usepackage{amssymb}
\usepackage{graphicx}
\usepackage{epsfig}
\usepackage{bm}
\usepackage{epstopdf}
\usepackage{array,subfig}

\begin{document}

\article{}{Theory of magnetic domains in uniaxial thin films}

\author{F. Virot$^1$, L. Favre$^1$, R. Hayn$^1$ and M. D. Kuz'min$^2$}

\address{$^1$IM2NP, Aix-Marseille Univ., Facult\'e St. J\'er\^ome, Case 142, F-13397 Marseille Cedex 20, France}

\address{$^2$IFW Dresden, P.O. Box 270 116, D-01171 Dresden, Germany}

\date{\today}

\begin{abstract}

For uniaxial easy axis films, properties of magnetic domains are usually described within the Kittel model, which assumes that domain walls are much thinner than the domains. In this work we present a simple model that includes a proper description of the magnetostatic energy of domains and domain walls and also takes into account the interaction between both surfaces of the film. Our model describes the behavior of domain and wall widths as a function of film thickness, and is especially well suited for the strong stripe phase. We prove the existence of a critical value of magneto-crystalline anisotropy above which stripe domains exist for any film thickness and justify our model by comparison with exact results. The model is in good agreement with experimental data for {\em hcp} cobalt.

\end{abstract}

\submitto{\JPD}

\maketitle

\section{Introduction}

Thin magnetic films with stripe domains and perpendicular magneto-crystalline anisotropy present a high 
scientific interest. They are model systems to understand the domain structure of ferromagnetic, as well 
as ferroelectric, thin films. Such materials are used for the fabrication of memories and spin injection 
devices. Considering a thin film with its bulk easy axis perpendicular to the surface, the magnetization 
direction in the domains can depend on the film thickness. Thick films behave as bulk material but for 
thin films, the competition between exchange, magnetostatic and magneto-crystalline energy can tip up 
the magnetization into the plane. A very well known theory of magnetic domain structure has been 
developed by \mbox{Kittel}~\cite{Kittel46}. In this model, the film magnetization is described by a 
square profile. Domain wall energy is allowed for the energy balance, but not for the magnetization 
profile, which determines the magnetostatic energy. Such an approximation is only valid when the domain 
width is much larger than the width of domain walls. Therefore, this model is bound to fail for very 
thin films, where the domain width vanishes. Furthermore, the magnetostatic interaction between the 
top and bottom surfaces is disregarded, which leads to wrong results unless the domain width is much 
less than the film thickness. The latter does not hold for very thin strongly anisotropic films.

Within the Kittel model, the width of stripe domains is proportional to the square root of the film
thickness. The model also predicts the existence of a critical film thickness below which the magnetization
direction flips from out-of-plane to in-plane direction. Later on, it was shown that for materials with 
strong magneto-crystalline anisotropy $K$ (in comparison to the square of the saturation 
magnetization $M_0$) the magnetostatic interaction energy between
top and bottom interface cannot be neglected. The ratio $2K/(\mu_0 M_0^2)$ (in SI units) 
is also known as the quality factor $Q$ and the magnetization direction remains 
perpendicular to the surface for all values of film thickness for $Q>1$. 
Numerous theoretical and experimental works have been devoted to such materials with 
strong perpendicular magnetic anisotropy \cite{Malek58,Kooy60,Marty97}. It has been shown that 
in such materials the domain width has a minimum as a function of film thickness, increasing
towards smaller or larger thicknesses. Materials with strong perpendicular anisotropy are also of 
technological importance in the field of magnetic multilayers (see \cite{Kiselev10} and references therein).  

Just above the critical film thickness, the magnetization still lies predominantly in the plane of the film,
deviations from the homogeneous in-plane orientation being very small. 
This permits an exact mathematical solution of the problem of stripe domain nucleation at the critical point 
\cite{Muller61}. Near the critical point the out-of-plane component of the magnetization remains much 
smaller than the saturation value $M_0$ and there is a strong variation of the 
magnetization direction across the film thickness. The corresponding phase is usually called 
the weak stripe phase. As the film thickness increases, the magnetization direction in the domains tends more 
and more towards the bulk easy axis, which is perpendicular to the film. Gradually, the weak 
stripe phase evolves towards the strong stripe phase, where the magnetization magnetization is predominantly
parallel (or antiparallel) to the bulk easy axis.

For a detailed analysis of the weak stripe phase a micromagnetic numerical analysis is indispensable 
(see \cite{Hubert98, Labrune01,Kisielewski04,Kisielewski07} and many 
other works). However, simple analytical models are helpful for getting a quick, albeit crude understanding 
of the stripe phases. These models restrict the magnetization direction to the vertical plane parallel to 
the stripes (i.e. they consider domain walls of Bloch type) but improve the Kittel model. One can distinguish 
between models with a linear domain wall profile~\cite{Zhao09} or others which use the Jacobi sine function 
to parametrize the magnetization profile~\cite{Sukstanskii97, Marty00}. Also a sinewave magnetization 
profile was proposed by Saito \cite{Saito64} to treat the weak stripe phase in an approximate manner. In the 
following we present a thorough analytical and numerical analysis of a simple but quite complete model for
the strong stripe phase. We adopt a sinewave magnetization profile in the wall and we take into account
the magnetostatic interaction between the top and bottom surfaces of the film. Our model applies
to hexagonal Co, for which a critical thickness between 25 nm \cite{hehnT,Hehn96} and  
40 nm  \cite{Brandenburg09} has been reported, as well as to materials with high magneto-crystalline 
anisotropy, such as FePd(001) or garnet films \cite{Marty97,Brandenburg09,Gemperle96}. Another possible 
application concerns ferromagnetic thin films of Mn$_5$Ge$_3$ that were recently synthesized 
\cite{Mendez08,Spiesser10}. Our model unifies previous results for materials with strong anisotropy ($Q>1$) 
\cite{Malek58,Kooy60,Marty97} and those with medium anisotropy ($\frac{1}{2} \lesssim Q<1$) 
\cite{Zhao09}, containing them as special cases. The 
explicit treatment of the domain wall is an important improvement with respect to the Kittel model
since it allows the determination of the in-plane remanent magnetization, which is important 
for the analysis of experimental data. We give numerical results for the dependence 
of domain and wall widths on the 
film thickness and calculate the critical thickness at which the magnetization switches from 
out-of-plane to in-plane direction if $Q<1$. We show that the critical thickness tends to zero as $Q \to 1$.
This behavior is identical in our model and in the exact solution \cite{Muller61}. In the case of strong
anisotropy ($Q>1$) stripe magnetic domains exist at any film thickness. 

Our theory considers a non-magnetistrictive film: the magneto-elastic energy is equal to zero. Moreover, 
the surface anisotropy is neglected. It is not suited for materials where the surface anisotropy 
plays an important role. We concentrate on the strong stripe phase since the details of the weak stripe phase are hard to capture in 
a simple model. In materials with high anisotropy the weak stripe phase does not appear at all and it will
be shown below that even in materials with medium anisotropy ($\frac{1}{2} \lesssim Q<1$) it is 
of rather limited importance. In the latter case the weak stripe phase occurs only in a narrow interval of 
thicknesses 
around the critical point. This region of the weak stripe phase is not correctly described in our model. 
On the other hand, even in the case of weak anisotropy ($Q \ll 1$), our model provides a correct qualitative 
description of the strong stripe phase, although it neglects such important phenomena as closure domains.

Our paper is organized as follows. After a short summary of known results 
(\mbox{Kittel}'s theory, sawtooth magnetization model, exact results for stripe phase nucleation) we present 
our model in Section~3. In Section~4 we show its analytical and numerical solution. The usefulness of our 
theory is demonstrated in Section~5 by way of theoretical analysis of published experimental data for 
the well-known {\em hcp} cobalt system. In Section~6 we compare our model with other approaches and expose
its strengths and limitations.

\section{Known results}

\subsection{Kittel's model}

\begin{figure}[ht]
\begin{center}
\includegraphics[width=.50\columnwidth]{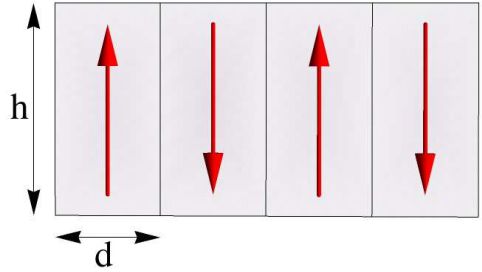}
\caption{\mbox{Kittel} model for ferromagnetic thin films with uniaxial anisotropy.}\label{fig:Kittel}
\end{center}
\end{figure}

This model was developed for ferromagnetic films with uniaxial anisotropy 
perpendicular to the film. It determines the width of magnetic stripe domains $d$ 
as a function of the film thickness $h$, and the critical thickness where the magnetization 
direction flips from out-of-plane to in-plane (see Fig.~\ref{fig:Kittel}). 
The total energy density (per unit volume) is given by:
\begin{equation}
E_{\rm total}=0.136 \mu_0 M_0^2 \frac{d}{h} + \frac{\sigma}{d}, \label{eK}
\end{equation}
where $M_0$ denotes the saturation magnetization and $\sigma$ the surface wall energy. 
The latter is obtained in a separate variational calculation for an isolated Bloch wall \cite{magn},
$\sigma=4\sqrt{AK}$, where $A$ is the exchange stiffness constant. The same calculation yields the wall
width, $\delta=\pi\sqrt{A/K}$. It is convenient to express all spatial dimensions in the units of
$\sqrt{A/K}$:

\begin{equation}
\tilde{h} = \frac{h}{\sqrt{A/K}}, \quad \tilde{d} = \frac{d}{\sqrt{A/K}},  \quad \tilde{\delta} = \frac{\delta}{\sqrt{A/K}}.
\label{reduced}
\end{equation}
Thus, Kittel's reduced wall width is a constant,

\begin{equation}
\tilde{\delta} = \pi.
\label{deltaK}
\end{equation}
The only essential material parameter in the model is the dimensionless quality ratio, $Q=2K/(\mu_0 M_0^2)$.
The first term of Eq.~(\ref{eK}) refers to the magnetostatic energy of a rectangular domain profile, 
neglecting the influences of finite wall width and film thickness. Therefore, strictly speaking, 
the \mbox{Kittel} model is only applicable when the following strong inequalities hold:
\begin{equation}
\delta \ll d \ll h,
\label{strong}
\end{equation}
or alternatively, $\tilde{\delta} \ll \tilde{d} \ll \tilde{h}$. 
The right-hand condition ensures the negligibility of the magnetostatic interaction between the top
and the bottom surfaces of the film. By minimizing the energy, one obtains the equilibrium domain width: 
\begin{equation}
\tilde{d} = 3.84 \sqrt{Q\tilde{h}},
\label{dopt}
\end{equation} 
and the minimum energy density:                                     
\begin{equation}
E^{\rm min}_{\rm total}=2\sqrt{\frac{0.136 \mu_0 M_0^2 \sigma}{h}} \; . \label{ev}
\end{equation}
As one can see, the minimum energy density $E^{\rm min}_{\rm total}$ tends to infinity as
$h \to 0$. Therefore, the multi-domain structure cannot be stable in that limit within the 
Kittel model. Indeed, for a uniform in-plane magnetization (perpendicular to the bulk easy axis), 
the magnetostatic and the exchange energies are equal to zero, and the anisotropy energy is maximal
and equal to $K$. The critical point is obtained by equating Eq.\ (\ref{ev}) to $K$.
Therefore, we can express the reduced critical thickness as follows:

\begin{equation}
\tilde{h}''_c = 4.35 Q^{-1}.
\label{eC} 
\end{equation}
Within the Kittel model, the domain width is identical to the half-period $d$ 
(see Fig.\ref{fig:Kittel}). 
Its value at the critical point is obtained by setting Eq.~(\ref{eC}) into~(\ref{dopt}), which
results in:

\begin{equation}
\tilde{d}''_c = 8 .
\label{dC}
\end{equation} 
These equations permit to obtain a qualitative evaluation of the critical quantities. But they 
fail in the neighborhood of the critical thickness, where domain and wall widths become close,
cf. Eqs. (\ref{deltaK}) and (\ref{dC}). Furthermore, by Eqs.~(\ref{eC}) and (\ref{dC}) in the critical 
region the ratio $h/d$ equals $0.544/Q = 0.272\mu_0 M_0^2/K$, i.e. also the second precondition 
of the Kittel model (\ref{strong}) is violated in the case of large anisotropy or small magnetization,
$Q\gtrsim1$. As a result, Kittel's model makes a wrong prediction for large $Q$: according to
Eq. (\ref{eC}), $\tilde{h}''_c$ decreases asymptotically, remaining nonzero at any finite $Q$.
In actual fact, the critical thickness should vanish at $Q=1$, as shown by rigorous calculations
\cite{Muller61}.

For that reason we decided to develop a model that would not break down in the strong-anisotropy case,
by allowing for the electrostatic interaction between the top and bottom surfaces of the film and 
taking into account the wall width explicitly. Within that improved  model we have obtained a good 
description of domain and wall width variation as a function of the film thickness (see Section 3).

\subsection{Sawtooth magnetization model}

The main goal of Kittel's model is the description of the strong stripe phase. Yet there is no direct transition
between strong stripes and planar magnetization. Rather, it proceeds via an intermediate weak stripe configuration. 
The latter can be described by sawtooth magnetization model  \cite{Saito64}. In this model the canting angle of the magnetization
out of the homogeneous in-plane direction, $\Theta(x)$, is assumed to be a saw-tooth curve with a maximum value 
of $\Theta_0$ and a minimum value of $-\Theta_0$, $\Theta_0 < \pi/2$. Between the extrema the variation is linear
and the half period is denoted by $d$. The sawtooth magnetization model is an approximate one, since it neglects the variation of 
the magnetization direction across the film thickness. The magnetostatic interaction between the top and bottom 
surfaces is not taken into account either. From the equation for the total energy below one can obtain the
expression for the critical thickness where the magnetization starts to flip out-of-plane by a small deflection angle. 
The first part of the total energy corresponds to the magnetostatic energy; $C_0$ is the leading coefficient 
of the Fourier series obtained from periodic profile of the deflection angle. 
The maximum angle $\Theta_0$ is a variational parameter, $0 < \Theta_0 < \pi/2$.

\begin{eqnarray}
\eqalign{E_{\rm{total}}=&\frac{\mu_0 M_0^2 C_0^2 d}{ 4 \pi  h}+\frac{4 \Theta_0^2 h}{d}+\frac{K}{2}\left(1+\frac{\sin 2 \Theta_0}{2 \Theta_0} \right) \\
\rm{with} \quad & C_0=\frac{2 \Theta_0 \cos \Theta_0 }{\pi^2 /4-\Theta_0^2}.\\}
\label{Esaito}
\end{eqnarray}
The critical thickness is expressed as : 

\begin{equation}
\tilde{h}'''_c = 27(2/\pi)^5 Q^{-1} = 2.82 Q^{-1}.
\label{ec:saito}
\end{equation}
The expression above has the same form as Kittel's critical thickness $\tilde{h}''_c$, Eq. (\ref{eC}), 
but with a smaller pre-factor. This opens up a possibility to interpret $\tilde{h}''_c$ and
$\tilde{h}'''_c$ as the upper and lower bounds of an interval where the weak stripe phase exists.
The latter is a transitional state between the homogeneously magnetized in-plane configuration, stable below
$\tilde{h}'''_c$, and Kittel's {\em strong} stripe domain structure, taking place above $\tilde{h}''_c$.
One observes the 'wrong' behavior of $\tilde{h}'''_c(Q)$ at $Q>1$, similar to that of $\tilde{h}''_c(Q)$.
Eq. (\ref{ec:saito}) also misbehaves at $Q \ll 1$. Rigorous calculations show (see Section 2.3) that
the true lower bound of the weak stripe phase does not diverge as $Q \to 0$.

\subsection{Exact description of stripe domain nucleation}

The exact solution of the nucleation problem was given in 1961 by Muller \cite{Muller61}. A modern
presentation, summarized here, can be found in the book of Hubert and Sch\"afer \cite{Hubert98}.
The theory was formulated for a thin magnetic film (thickness $h$, magnetization $M_0$) with 
uniaxial anisotropy perpendicular to the film (anisotropy constant $K$). The energy expression contains the 
magnetostatic energy, the exchange (stiffness $A$) and anisotropy contributions; the theory imposes no 
restrictions on the magnetization direction. Below the critical thickness $h_c$ the magnetostatic energy forces
the magnetization into the plane of the film. At $h=h_c$ there is an instability and the weak stripe phase emerges. 
In the neighbourhood of the critical point, deviations from a homogeneous in-plane orientation of
magnetization are small, which allows to linearize the system of micromagnetic equations and find the exact
analytical form of the instability mode. The reduced critical thickness $\tilde{h}_c$ is a universal function 
of the quality ratio $Q$, see Fig. \ref{fig:Validity} (solid line). (The quantity plotted in the original drawing,
Fig. 3.109{\em a} of Ref.\ \cite{Hubert98}, is a factor $2\pi$ smaller.)
At the same time, the critical half-period $d_c$ can be found, as well as the distribution of magnetization 
directions in the critical mode. In the weak-anisotropy limit, $Q \to 0$, both $\tilde{h}_c$ and $\tilde{d}_c$ 
tend to the same finite value,

\begin{equation}
 \tilde{h}_c \to \tilde{d}_c \to 2 \pi.
 \label{ecM}
 \end{equation}
As $Q$ increases, the critical thickness $h_c$ decreases and vanishes at $Q=1$, whereas $d_c$ diverges at that
point. The exact results will be used later on to judge the validity of the sinewave wall model.

\section{Sinewave wall model (SWM)}

\begin{figure}[ht]
\begin{center}
\includegraphics[width=1.00\columnwidth]{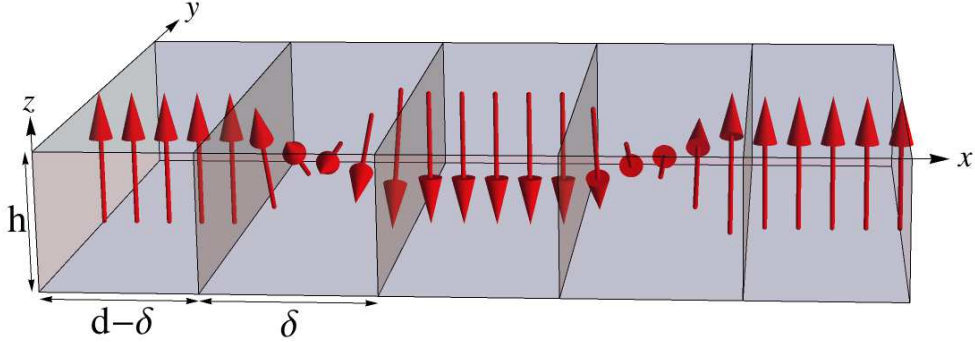}
\caption{Sketch of periodic stripe domains in a magnetic thin film with domain walls of finite thickness.}\label{fig:CLW}
\end{center}
\end{figure}

\begin{figure}[ht]
\begin{center}
\includegraphics[width=1.00\columnwidth]{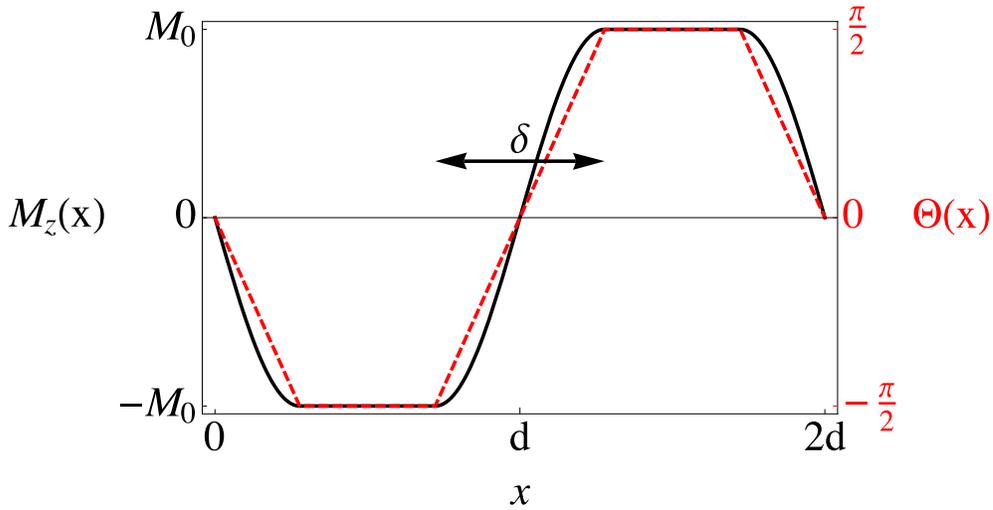}
\caption{(color online) One full period of the domain structure of Fig.~\ref{fig:CLW}. $M_{z}(x)$ (solid line) 
denotes the $z$ component of the film magnetization, and $\Theta (x)$ is the angle between the magnetization 
direction and the film surface (dashed line). The domain wall width is denoted by $\delta$.}\label{fig:Mz}
\end{center}
\end{figure}

We propose a model stripe domain structure as shown in Figs. \ref{fig:CLW} and \ref{fig:Mz}.
Like in Kittel's model, the half-period is denoted by $d$, but it now includes a domain wall of width 
$\delta$ (see Fig.~\ref{fig:CLW}). The inner domains, with constant magnetization equal to 
$\pm M_0$, have a width of $d-\delta$. 
The walls separating different domains are assumed to be of Bloch type, with linear variation 
of the angle $\Theta(x)$ between the magnetization direction and the $y$-axis. This results 
in a magnetization profile of sinewave form, which is a good approximation of the
profile obtained by the variational method. The schematic representation in Fig.~\ref{fig:CLW} 
shows that we consider parallel stripe domains infinite in the $y$ direction. In the $x$ direction, 
the periodic magnetization profile is as shown in Fig.~\ref{fig:Mz}. The $z$ dimension is restricted
to the film thickness $h$.

To calculate the magnetostatic energy, we use an analogy with the electrostatic field calculation
for alternating, positively and negatively, charged stripes (see Landau-Lifschitz ~\cite{Landau56}). 
We consider the realistic magnetization profile of Fig.~\ref{fig:Mz} (as opposed to a simple 
rectangular meander) as well as the magnetostatic interaction between the top and bottom surfaces
of the film.

The magnetostatic energy density (per volume) can be written as:

\begin{equation}
E_{\rm{mag}}=\frac{\mu_0  d }{\pi  h}\sum _{k=1}^{\infty } \frac{|C_k|^2}{k} \left[ 1-\exp{\left(-\pi k \frac{h}{d}\right)}\right] \label{mag}
\end{equation}

\begin{equation}
\mbox{with} \quad C_k=\frac{2 M_0}{k \pi  \left[1-k^2(\delta/d)^2\right]}\cos\left( \frac{k \pi \delta}{2 d}\right) \label{Cf}
\end{equation}
where $k$ is an odd number, $C_k$ is the \mbox{Fourier} coefficient of $M_{z}(x)$ (see Fig. \ref{fig:Mz}). 
The domain structure is determined by the interplay between magnetostatic, exchange and anisotropy energies. 
The latter two contributions are as follows:

\begin{eqnarray} 
\eqalign{E_{\rm{exch}}&=\frac{1}{2d}\int _{0 }^{2d}A \left(\frac{\rm{d}\Theta }{{\rm d} x}\right)^2 {\rm d}x \\ \label{fech}
&=\frac{\pi ^2}{d \delta }A\\} 
\end{eqnarray}

\begin{eqnarray}
\eqalign{E_{\rm{aniso}}&=\frac{1}{2 d}\int _{0 }^{2 d}K\cos^2[\Theta] {\rm d}x\\ \label{faniso}
&=\frac{ \delta }{2 d}K}
\end{eqnarray}
The half-period $d$ and the wall width $\delta$ are then calculated by minimizing the total energy,

\begin{equation}
E_{\rm{total}}=E_{\rm{mag}}+E_{\rm{exch}}+E_{\rm{aniso}}\label{eq-et}.
\end{equation}
Since it is not possible to solve analytically the two variational equations $\partial _dE_{\rm{total}}=0$, 
$\partial _\delta E_{\rm{total}}=0$ in the general case, we use a Nelder-Mead numerical method. The results 
are presented in the next Section.

\section{Analytical and numerical results}

\subsection{Large film thickness}

Before presenting the numerical solution in the general case let us discuss some specific situations. 
For sufficiently thick films the strong inequalities (\ref{strong}) are fulfilled and the expression 
for the \mbox{Fourier} coefficients, Eq.~(\ref{Cf}), becomes: 

\begin{equation}
C_k=\frac{2 M_0}{k \pi}.
\end{equation}
Then Kittel's form is recovered for the magnetostatic energy density:
\begin{equation}
E_{\rm mag}=\frac{4 \mu_0 M^2_0 d}{\pi^3 h}\sum _{k=1}^{\infty }k^{-3}= 0.136 \mu_0 M^2_0 \frac{d}{h}. 
\end{equation}
The wall energy associated with the above magnetostatic expression is a minimum for the following reduced 
wall width:
\begin{equation}
\tilde{\delta}_{\infty} = \pi \sqrt{2}. 
\label{delta_inf}
\end{equation}

The corresponding surface wall energy, $\sigma=\pi\sqrt{2}\sqrt{AK}$,
is slightly (11\%) larger than the variational result, $4\sqrt{AK}$, because of the imposed sinewave
wall profile of $M_z(x)$.
In the thick-film limit the dependence of the domain width on film thickness can be expressed analytically,
 \begin{equation}
 \tilde{d} = 4.05 \sqrt{Q\tilde{h}},
 \label{eq-kit}
 \end{equation}
and is very similar to the result of the \mbox{Kittel} model, Eq.~(\ref{dopt}).

\subsection{Critical thickness and critical anisotropy}

\begin{figure}[!ht]
\begin{center}
\includegraphics[width=1.00\columnwidth]{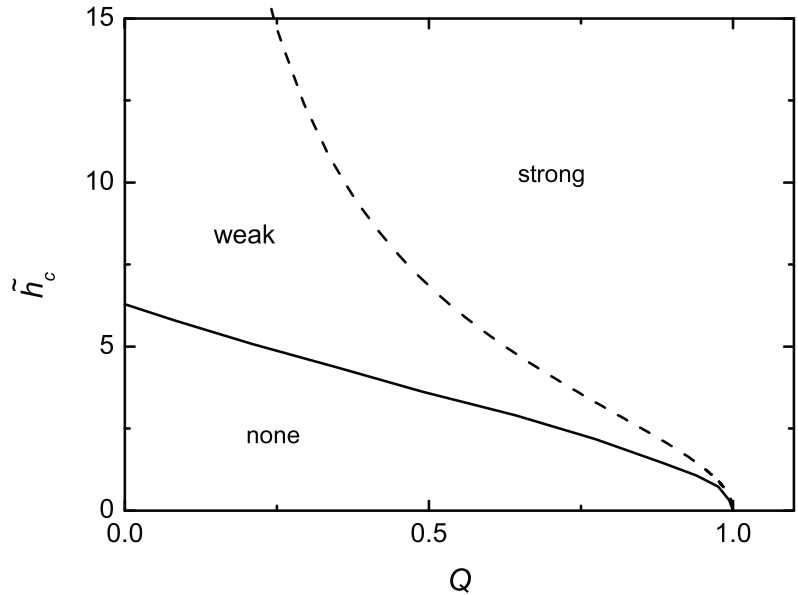}
\caption{Phase diagram of thin magnetic films in zero magnetic field. The solid line is the critical
thickness $\tilde{h}_c$ obtained in the exact approach \cite{Muller61,Hubert98}, it separates the
homogeneous state without domains from the weak stripe phase. The dashed line is the reduced
critical thickness $\tilde{h}'_c$ of SWM, computed using Eqs. (\ref{param_Q}) and (\ref{param_ec}).
This is a cross-over line between the weak and the strong stripe phases.}\label{fig:Validity}
\end{center}
\end{figure}

As the film thickness is reduced, the domain width $d$ decreases and the wall width $\delta$ increases. 
Just above the critical thickness, the magnetization profile is purely sinewave ($\delta \to d$), 
forming an unstable spiral magnetic configuration. In this limit Eq.~(\ref{mag}) can be simplified: 
the sum disappears since only the leading term of the \mbox{Fourier} series, with $k=1$, survives. 
The magnetostatic energy becomes: 
\begin{equation}
E_{\rm mag}=\frac{\mu_0 M^2_0 d}{4 \pi h} \left[ 1- \exp{\left(-\pi\frac{h}{d}\right)} \right]    \label{Emagec}
\end{equation}  
and anisotropy plus exchange terms (\ref{fech},\ref{faniso}), can be written as: 
\begin{equation}
E_{\rm wall}=\frac{K}{2}+\frac{\pi^2 A}{d^2}.
\end{equation}
Minimizing the total energy, $E_{\rm mag}+E_{\rm wall}$, with respect to $d$ and equating the result to $K$ 
yields the critical thickness $h'_c$: for this film thickness the total energies of the stripe domain 
structure and of a mono-domain state with in-plane magnetization are equal. The corresponding value of the
half-period is $d'_c$. Both $h'_c$ and $d'_c$ depend on the quality ratio,
$Q=2K/(\mu_0 M_0^2)$. This dependence can be presented in parametric form by introducing an auxiliary
quantity, $t=\pi h'_c/d'_c$. One then finds
\begin{equation}
Q = \frac{3}{2} \left[ \frac{1}{t} -\left( \frac{1}{3} +\frac{1}{t} \right) e^{-t} \right],
\label{param_Q}
\end{equation}
\begin{equation}
\tilde{h}'_c = \sqrt{6} \, t \, \sqrt{\frac{1-(1+t/3)e^{-t}}{1-(1+t)e^{-t}}},
\label{param_ec}
\end{equation}
and
\begin{equation}
\tilde{d}'_c = \pi t^{-1} \tilde{h}'_c.
\label{param_dc}
\end{equation}
The parameter $t$ runs from zero to infinity. The resulting $\tilde{h}'_c$-vs-$Q$ dependence is displayed
in Fig. \ref{fig:Validity} (dashed curve). It can be regarded as a borderline between the strong- and the
weak-stripe phases. The solid line in Fig. \ref{fig:Validity} presents the exact result 
of Muller's theory \cite{Muller61,Hubert98}, $\tilde{h}_c(Q)$. This line separates the region of weak stripes from 
the area of homogeneous in-plane magnetization, as observed in very thin films with $Q<1$.

For $Q$ small, $Q \to 0$, $t \to \infty$, the critical thickness $\tilde{h}'_c$ diverges,
\begin{equation}
\tilde{h}'_c \approx 3\sqrt{\textstyle\frac{3}{2}} Q^{-1} \approx 3.67 Q^{-1}, 
\label{ec_div}
\end{equation}
whereas
\begin{equation}
\tilde{d}'_c \to \pi \sqrt{6}.
\label{dc_lim}
\end{equation}
The opposite limiting case is $t \to 0$, $Q \to 1$. In that limit $\tilde{h}'_c$ tends to zero,
\begin{equation}
\tilde{h}'_c = 4\sqrt{2(1-Q)} \; ,
\label{m0crit}
\end{equation}
while $\tilde{d}'_c$ diverges,
\begin{equation}
\tilde{d}'_c = \pi \sqrt{ \frac{2}{1-Q} }.
\label{dc_div}
\end{equation}
For strong anisotropy, $Q>1$ or $K>\frac{1}{2}\mu_0 M_0^2$, there is no physically meaningful $h'_c$ 
or $d'_c$ and,
without a magnetic field, the mono-domain structure with in-plane magnetization is unstable for any $h$. 
We find it remarkable that our model reproduces the exact value for the critical quality ratio, $Q=1$
\cite{Muller61,Hubert98}.
The Kittel model is limited to systems with small quality ratios. However, in many strongly anisotropic 
materials $Q$ is large and no critical thickness is observed \cite{Marty97,Gemperle96,Yang11}.

\subsection{Numerical results}

\begin{figure}[!ht]
\begin{center}
\includegraphics[width=1.00\columnwidth]{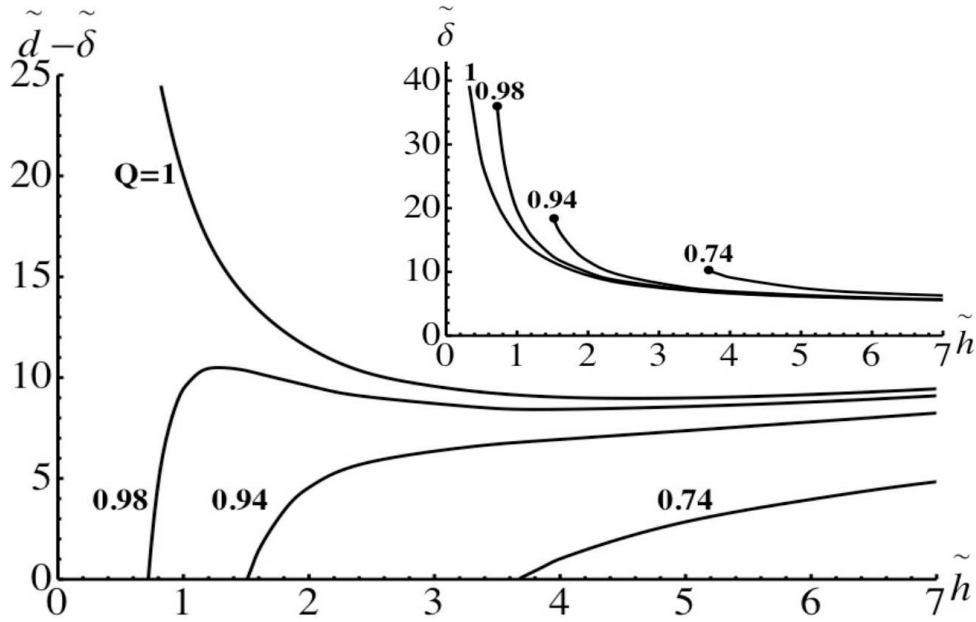}
\caption{Inner domain width $\tilde{d}-\tilde{\delta}$ and wall width $\tilde{\delta}$ as a function 
of film thickness $\tilde{h}$. All figures are presented in reduced coordinates (\ref{reduced}). 
Bold numbers correspond to the value of $Q$.}\label{fig:Wdimless}
\end{center}
\end{figure}

Away from the critical region, the total energy (\ref{eq-et}) has to be minimized with respect to $d$ and 
$\delta$. In general, for arbitrary $M_0$, $K$, $A$, and $h$, the result cannot be 
expressed analytically and we have to resort to a numerical procedure. The dependence on $A$ and $K$ can still 
be taken into account by changing over to the dimensionless variables (\ref{reduced}).
Then we are left with two quantities $\tilde{\delta}$ and $\tilde{d}-\tilde{\delta}$ versus $\tilde{h}$ with only 
one parameter $Q$. In Fig. \ref{fig:Wdimless} we prefer to plot the difference $\tilde{d}-\tilde{\delta}$ rather 
than $\tilde{d}$ 
since $\tilde{d}-\tilde{\delta}$ can be regarded as an order parameter. The numerical curves confirm all 
analytical expressions, in particular, those describing the asymptotic behavior at $\tilde{h} \gg 1$, 
Eqs. (\ref{delta_inf}) and (\ref{eq-kit}). Thus, $\tilde{\delta}(\tilde{h})$ can be seen to tend to a universal 
limit. 

One observes in Fig. \ref{fig:Wdimless} two distinct regimes, for $Q<1$ and for $Q>1$. If $Q>1$, the stripe
domain structure is stable at any $\tilde{h}$. The half-period of the structure, $d(h)$ or $\tilde{d}(\tilde{h})$,
has a minimum at a certain finite thickness, increasing towards smaller and larger $\tilde{h}$. A prominent
feature of the curves with $Q<1$ is the presence of a critical thickness $\tilde{h}'_c$ where the
width of the inner domains, $\tilde{d}-\tilde{\delta}$, vanishes. At that point the sample contains nothing
but domain walls, the magnetic structure being a spiral of period $2d_c = 2\delta_c$. Obviously, the wall width
cannot be neglected, especially at $Q\approx 1$.

\subsection{In-plane remanent magnetization}

In a magnetic field directed in the plane of the film the stripe domains arrange themselves parallel 
to the external field (see Refs. \cite{hehnT,Hehn96}). Each domain wall has a magnetic 
moment component parallel to the in-plane field direction, all of them being directed in the same 
sense for a small but finite external field. These contributions sum up to a 
finite in-plane remanent magnetization with a corresponding hysteresis,
as observed in thin $hcp$ Co or Mn$_5$Ge$_3$ films \cite{Spiesser10,Hehn96}. 
This remanent in-plane magnetization (when the external field tends to zero) is easy to calculate 
in our model. The sinewave magnetization profile in the domain wall leads to the following expression:

\begin{equation}
\frac{M_r}{M_0}=\frac{1}{d}\int_0^{\delta}\sin{\frac{\pi x}{\delta}} \; {\rm d} x
=\frac{2\delta}{\pi d}\label{mr} \; .
\end{equation}

At the critical thickness, the remanent magnetization is equal to 
$2/\pi\simeq 63.7$ \% of the saturation magnetization. 
Just below this point we would expect the weak stripe phase \cite{Saito64} in a narrow interval of
thicknesses before the transition into the planar mono-domain state. In the weak stripe phase
the maximum $z$-component of the magnetization is less than $M_0$, like in sawtooth magnetization model
\cite{Saito64}. However, the weak stripe phase is beyond the scope of our model.

\section{Hexagonal Co}

\begin{figure}[!ht]
\begin{center}
\includegraphics[width=1.00\columnwidth]{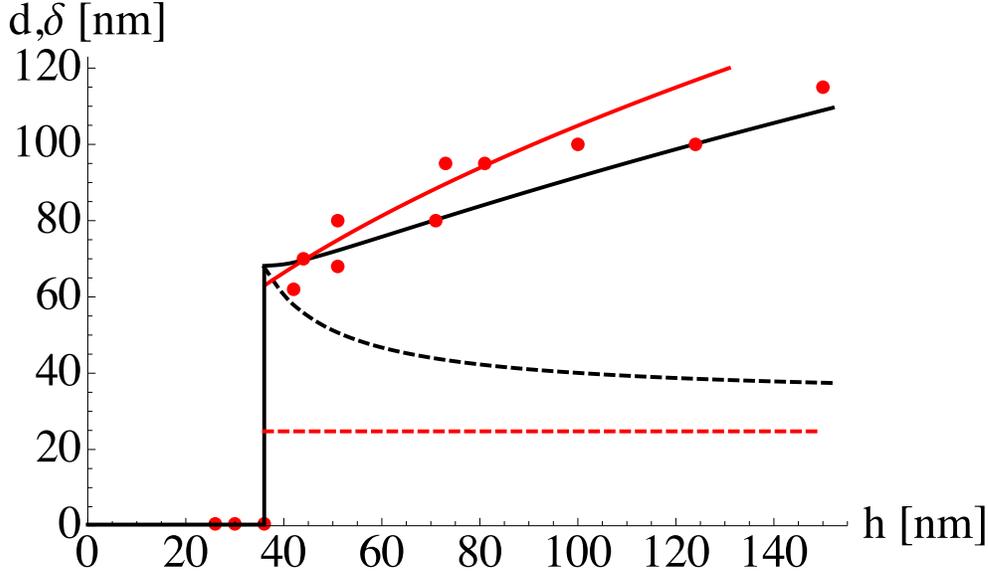}
\caption{Half-period $d$ (full line) and wall width $\delta$ (dashed line) as functions
of thickness for Co films. Red points are experimental $d$ values from Ref. \cite{Brandenburg09}. 
Red and black curves are fits to the Kittel model and SWM, respectively.}\label{fig:Period}
\end{center}
\end{figure}

Cobalt hexagonal thin films have been intensively studied. Below 
a critical thickness, such films exhibit planar magnetization. Recent experimental 
publications~\cite{Brandenburg09,Hehn96,Donzelli08} report critical thicknesses
between 25 and 50 nanometers. Since the study of Brandenburg {\em et al.}~\cite{Brandenburg09} 
is the most complete one, SWM has been tested using those data, especially 
the half-period as a function of film thickness, shown in Fig.~\ref{fig:Period}.

SWM has three parameters to characterize different materials: the magneto-crystalline 
anisotropy constant $K$, the saturation magnetization $M_0$ and the exchange stiffness constant $A$.
The saturation magnetization was fixed to the experimental value of $M_0=1.43$ MA/m, while the anisotropy 
as well as the exchange constant were chosen as fitting parameters. The best fit was obtained with 
$K=820$ kJ/m$^3$, which is rather close to the bulk anisotropy constant (see Tab.~\ref{tab:fit}), 
and $A=45$ pJ/m, which exceeds the value deduced from inelastic neutron scattering experiments
\cite{Vaz08}. However, it should be taken into account that $A$ is not known accurately,
since relating neutron scattering data with our continuum model involves several approximations.

\begin{table}[ht]
\begin{center}
\renewcommand{\arraystretch}{1.2}
\begin{tabular}{|l||c|c|c|}
\hline 
cobalt & Kittel & SWM & measurements \\ \hline \hline
$K$ [kJ/m$^3$] & 1200 & 820 & 820$^a$, 920$^b$  \\ \hline
$A$ [pJ/m] &76 & 45 & 28$^c$\\ \hline 
$M_0$ [MA/m]& 1.43$^*$ & 1.43$^*$ & 1.43$^d$ \\ \hline  \hline 
$h_{\rm{c}} $ [nm] & 36 & 36 & 36 $\pm3$ \\ \hline 
$d_{\rm{c}}$ [nm] & 63 & 67 & 63 $\pm5$ \\ \hline
\end{tabular}
\caption{Summary of important physical constants obtained for cobalt. The asterisks indicate fixed parameters. 
Experimental data are extracted from: $^a$ Ref. \cite{Paige84}, $^b$ Ref. \cite{hehnT}, 
$^c$ Ref. \cite{Vaz08}, $^d$ Ref. \cite{Brandenburg09,Donnet95}}
\label{tab:fit}
\end{center}
\end{table} 

For comparison, we also fitted the experimental data to the Kittel model (see Fig.~\ref{fig:Period}),
leading to the parameter values presented in Table~\ref{tab:fit}. The saturation magnetization 
was fixed to the same value, but the anisotropy and exchange constants deviate more strongly from the
experiment than those obtained in the SWM fit. This suggests that the Kittel model is less realistic
than SWM. We also observe important differences between both models near the critical thickness. The curvature 
of $d(h)$ near $h_c$ is different in the two models. In SWM, for $h$ between 40 and 80 nm the wall is wider
than the inner domain, whereas in the Kittel model the wall width remains constant and always inferior to
the inner domain width.

The quality factor for Co is $Q=0.64$, which is less than one. We expect that for materials with stronger
anisotropy, i.e. with larger $Q$, the difference between SWM and the Kittel model will be even
more significant.

\section{Comparison}

Our model is constructed in such a way that it applies at any $Q$, including the strong-anisotropy case, $Q>1$,
where there is no critical thickness.

The most interesting region is the one of medium anisotropy, $\frac{1}{2} \lesssim Q<1$. To 
illustrate that region, we chose hexagonal cobalt as an example. With the parameters of Table 1 (SWM)
we evaluated the critical thickness in SWM, $h'_c=36$ nm. This agrees with the experiment
of Brandenburg et al. \cite{Brandenburg09}, who find a critical thickness of about 40 nm. However, in Muller's
exact theory \cite{Muller61} the critical thickness comes out much smaller, $h_c=23$ nm. One has to take into
account, though, that stripe domains were observed by other groups in Co films as thin as 25 nm and that
weak stripes are certainly hard to see. Our interpretation is that the interval between Muller's $h_c=23$ nm
and the SWM $h'_c=36$ nm is a region of the weak stripe phase.

\begin{figure}[!ht]
\begin{center}
\includegraphics[width=1.00\columnwidth]{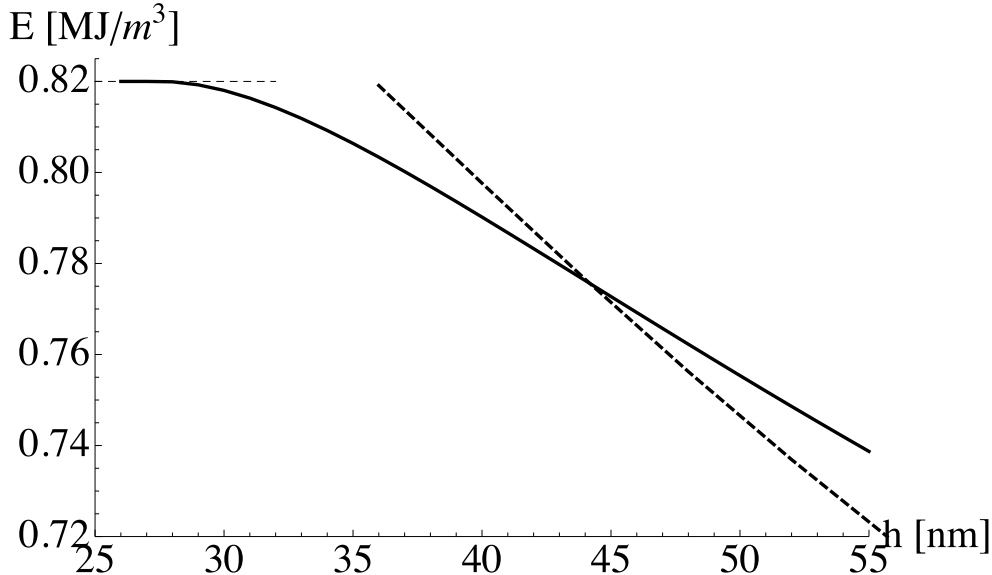}
\caption{Comparison of the total energy of the weak stripe phase (sawtooth magnetization model, solid line) with that one of the 
strong stripe phase (SWM, dashed line).}\label{fig:Saito2}
\end{center}
\end{figure}

In the case of cobalt, such an interpretation is supported by a calculation of the weak stripe phase within 
the sawtooth magnetization model. The two variational parameters are the maximum canting angle $\Theta_0$ and the half-period $d$.
For better precision, we included the complete Fourier series as well as the magnetostatic interaction between 
the top and bottom surfaces. The numerical results are presented in Fig. \ref{fig:Saito2}. 
The weak stripe phase sets in at $h=28$ nm and has a lower energy than the strong stripe phase up to 44 nm. 
Between 28 nm and 44 nm the maximum canting angle increases from zero to 76 degrees. So, in the case of hexagonal 
cobalt, we would expect the existence of the weak stripe phase between 23 and about 44 nm. At larger thicknesses
the strong stripe phase prevails and SWM is more appropriate than the sawtooth magnetization model. 

Let us now briefly discuss the region of very small anisotropy, $Q \ll 1$.
It allows mathematical simplifications, as already discussed in Section 4. There, it was shown
that the range of thicknesses where the weak stripe phase could be expected becomes wider according as $Q$
decreases, i.e. as the magnetic anisotropy weakens. For very weak anisotropy we also expect other complications,
such as the presence of closure domains. We believe our model can still be applied if the layer of the closure 
domains is much thinner than the film as a whole.

\section{Conclusion}

We present a model (SWM) for the strong stripe phase in magnetic thin films with perpendicular anisotropy,
which improves the well-known Kittel model in two respects. Firstly, our model includes the magnetostatic 
interaction between the top and bottom surfaces of the film. This is important for materials with strong 
magnetocrystalline anisotropy. Secondly, the domain walls are treated explicitly, which improves the numerical
accuracy and allows the calculation of the in-plane remanent magnetization. Our model is simple, but general enough 
to permit a thorough analysis of the strong stripe phase. It is especially suited to interpret experimental data. 
We demonstrate the existence of a critical anisotropy above which stripe domains exist at any film thickness. 
SWM reproduces the exact threshold value, $Q=1$. We derive a number of analytical results facilitatng the 
estimation of important parameters. The numerical results obtained for different quality factors $Q$ show the 
evolution of domain and wall widths as functions of film thickness.

With the new model we are able to correctly describe the behavior of magnetic domains in cobalt thin films. 
It allows to explain the experimental observations of Brandenburg {\em et al} \cite{Brandenburg09} 
with more realistic fitting parameters than using the Kittel model. By comparing the exact critical thickness
for stripe nucleation $h_c$ with our critical thickness $h'_c$, which corresponds to the on-set 
of strong stripes, we are able to estimate the range of thicknesses where weak stripes are expected. 
We show that for Co this interval is rather narrow, 10 to 15 nm, which validates our model. We should remark 
that the currently available experimental data for Co do not suffice for locating the thickness range 
of the weak stripe phase accurately. We find it important that our model takes the wall width
into account in an adequate way. In the neighborhood of the critical thickness, domain and wall width are
of comparable size. Thanks to its universality, our model can be applied to other types of ferromagnetic 
films, e.g., FePd or Mn$_5$Ge$_3$. 

\ack{This work was supported by PICS project (No. 4767) and by ANR-project MNGE-SPIN. The authors are grateful 
to A.N. Bogdanov for illuminating discussions.}

\end{document}